\documentclass[preprint,prd,showpacs]{revtex4}
\newcommand{\teta}{\rlap{\lower2ex\hbox{$\,\tilde{}$}}\eta{}}

\newcommand{\Vec}[1]{{\mathbf{#1}}}

\begin{document}

\title{Bounds on Stringy Quantum Gravity from Low Energy Existing Data}

\author{Daniel Sudarsky}
\affiliation{Center  for Gravitational Physics and
    Geometry,
Department of Physics, Penn State University, University park, PA
16802, USA}
\altaffiliation[On leave of absence from]{ Instituto de Ciencias Nucleares,
  Universidad 
Nacional Aut\'onoma de M\'exico, A. Postal 70-543, M\'exico D.F.
04510, M\'exico}
\author{Luis Urrutia}
\affiliation{ Instituto de Ciencias Nucleares,
Universidad Nacional Aut\'onoma de M\'exico, A. Postal 70-543,
M\'exico D.F. 04510, M\'exico}
\author{H\'ector Vucetich}
\affiliation{ Instituto F\'\i sica,
 Nacional Aut\'onoma de M\'exico,
A. Postal 70-543, M\'exico D.F. 04510, M\'exico}
\altaffiliation[On leave of absence from]{ Observatorio Astron\'omico,
Universidad Nacional de La Plata, La Plata, Argentina}

\begin{abstract}
We show that existing low energy experiments, searching for the
breaking of local Lorentz invariance, set bounds upon string
theory inspired quantum gravity models that induce corrections to
the propagation of fields. Using the standard Observer Lorentz
transformations in the D-particle recoil model we find $M \geq
1.2 \times 10^5 M_P$ and $v \leq 2\times 10^{-27}\, c$ for the mass
and recoil speed of the D-particle, respectively. These bounds are
$\sim 10^8$ times stronger than the latest astrophysical bounds. These
results indicate that the stringy scenario for modified dispersion
relations is as vulnerable to these types of tests as the loop
quantum gravity schemes.
\end{abstract}


\maketitle

\section{Introduction}
\label{Intro}

We have recently analyzed the bounds \cite{Sudarsky:2002ue} that can be
extracted form direct searches of violations of Lorentz invariance
\cite{Kostelecky:1989zi,Kostelecky:1989jw,Kostelecky:1991ak,%
Kostelecky:1996qk,Colladay:1997iz,Colladay:1998fq,Kostelecky:2001xz,%
Bluhm:2001yy} that seem inevitable in the scenarios that are
used to predict changes in the dispersion relations of high energy
particles due to quantum gravitational effects 
\cite{Amelino-Camelia:1998gz,Ahluwalia:1999aj,%
Amelino-Camelia:1999zc}.
Our original analysis dealt specifically with the Loop Quantum
Gravity scenarios\cite{Gambini:1998it,Alfaro:2002xz,Alfaro:1999wd,%
Alfaro:2001rb,Urrutia:2002tr}, however as suggested in Ref.
\cite{Sudarsky:2002ue}, we expected similar results to apply to any
such scheme and in particular to the String Theory inspired models
\cite{Ellis:1999uh,Ellis:1999jf,Ellis:1999sd,Ellis:2000sx,%
Gravanis:2002ii,Ellis:1999sf,Ellis:2002bu}.

The main point is that  when a theory predicts that photons
propagate with an energy-dependent velocity $v(E)$ rather than
with the universal speed of light $c$,  then such theory is
automatically implying  a breakdown of Lorentz invariance, either
 at the fundamental or  at the spontaneous level, since such
statement can be  valid at best in one specific inertial
frame. This selects a preferred frame of reference, where the
particular form of the corrected equations of motion are valid,
and one should then be able to detect the laboratory velocity with
respect to that frame. Physics present us with  a rather unique
canonical choice for that ``preferred inertial frame'': that one
where the Cosmic Microwave Background (CMB) looks isotropic. Our
velocity $\Vec{w}$ with respect to that frame has already been
determined to be $w/c\approx 1.23 \times 10^{-3}$ by the
measurement of the dipole term in the CMB by COBE, for instance
\cite{1996ApJ...470...38L}.  Thus, it follows that the quantum gravity
corrections to the corresponding particle field theory (photons,
{fermions} and others) should contain $\Vec{w}$-dependent terms
when described in our laboratory reference frame. These would lead
to a breakdown of isotropy which would show up in high precision
tests of rotational symmetry, that have been carried out using
atomic and nuclear systems such as those described in
\cite{HRBL60,Drever61,Prestage85,Lamoreaux:1986xz,Lamoreaux89,%
Chupp89,Berglund95,Bear:2000cd,Phillips:2000dr}.

The method of analysis  is the same used in  our previous work and
can be thought to correspond to the application of the general
framework described in the works  of
Ref. \cite{Colladay:1997iz,Colladay:1998fq}, to the 
specific scenario in question.

The specific String Theory inspired scenario that has been
considered as leading to a quantum gravity induced change in the
dispersion relation of high energy particles is the so called
Liouville approach to Non-critical String Theory. In this scheme
our universe is identified with a 4 dimensional D-brane, which
will naturally contains a type of topological defect called a
D-particle. These will interact with ordinary particles,
represented by strings in the corresponding mode, by elastic
scattering, which would produce a recoil of the D-particle. This
recoil will induce a local disturbance of the space-time geometry
which will in turn affect the propagation of the particle
\cite{Ellis:1999uh,Ellis:1999jf,Ellis:1999sd,Ellis:2000sx,%
Gravanis:2002ii,Ellis:1999sf,Ellis:2002bu}.
 In the following, we shall use
``brane'' and ``D-particle'' as synonymous terms.

 We take this part of the analysis
directly from Ref. \cite{Ellis:2002bu}, starting with their modified
Dirac equation
\begin{eqnarray}
\left[{\gamma}^{\mu }(i\partial _{\mu }-eA_{\mu })-v^{i}{\gamma}
^{0}(i\partial _{i}-eA_{i})-m\right] \Psi &=&0. \label{DIREQ}
\end{eqnarray}
Here $\gamma^\mu$ are the standard flat-space gamma matrices and
$v^i$ is the D-particle recoil velocity in the CMB frame, where it
is initially at rest. The above modified Dirac equation preserves
gauge invariance due to the standard minimal coupling. In the
following we set $A_\mu=0$. We will be concerned with a
perturbative expansion in the small parameter $v^{i}$.
Note that such modification of Dirac's equation,
\begin{bf} having a privileged
constant vector $v^i$, 
\end{bf}
might arise in 
different types of quantum gravity models.

Already at this level  one might consider two alternatives in
addressing the model based in Eq.(\ref{DIREQ}):

(i) the first one, proposed in Ref. \cite{Ellis:2002bu}, 
argues that Eq.(\ref{DIREQ}) is form invariant, while maintaining
the meaning of $v^i$ as the brane recoil velocity produced by the
particle-brane collision in all references frames. This property,
suggested by the underlying string theory basis of (\ref{DIREQ}),
is realized by demanding that $v^i$ is embedded in a tetrad
$e_a{}^\mu$ with components;
\begin{equation}
e_0{}^0=1, \quad e_0{}^i=v^i, \quad e_i{}^0={\Vec 0}, \quad
e_i{}^j=\delta^j_i, \label{tetrad}
\end{equation}
 in each reference frame.  The above leads to a modified
Minkowski metric $G_{\mu\nu}$ producing the following invariant in
momentum space
\begin{equation}
 p^\mu\,G_{\mu\nu}\,p^\nu=E^2-2E{\Vec v}\cdot{\Vec p}-{\Vec
p}{}^2, \quad p^{\mu}=(E, {\Vec p}), \quad |{\Vec v}|\approx
\frac{\mid \Vec{p} \mid }{M}. \label{metric} 
\end{equation}
 where $M$ is
the D-particle's mass.

 As stated in Ref. \cite{Ellis:1999sf}, the set of transformations leaving
invariant either (\ref{tetrad}) or (\ref{metric}) is a subgroup of the
standard Lorentz group. Application of this idea then would mean, at
least, to abandon the use of the standard Lorentz transformations to
connect inertial reference frames.  Since one is violating Lorentz
covariance, this is a possible alternative which remains to be
explored in full detail.  However, we must point out that it is
thus unclear what would be the transformation properties of the Dirac
wave function $\Psi $ and of the equation (\ref{DIREQ}), and in which
frame should we consider such description to be valid.

(ii) Here we take a more conservative approach by maintaining the
full Lorentz covariance relating any two inertial frames (Observer
Lorentz covariance), according to the standard model extension of
Kosteleck\'y et. al
\cite{Kostelecky:1989zi,Kostelecky:1989jw,Kostelecky:1991ak,%
Kostelecky:1996qk,Colladay:1997iz,Colladay:1998fq,Kostelecky:2001xz,%
Bluhm:2001yy}. Then we expect the metric
(\ref{metric}) to transform as a second-rank tensor, while the brane
recoil velocity is embedded in the four vector $V^\mu$ defined as a
momentum transfer. To order $1/M$ we have $V^\mu=(0,{\Vec v })$ in the
CMB frame. As expected, the collision calculation of the brane recoil
velocity in each reference frame is consistent with the standard
Lorentz transformation properties of the four-vector $V^\mu$. Also we
take a perturbative approach where the brane recoil velocity (which is
of order $|{\Vec p}|/M$) is calculated using the uncorrected
energy-momentum dispersion relations. We have  checked that the
use of the modified dispersion relations arising from
Eq.(\ref{metric})  does not change the results to that order.

The paper is organized as follows: In section \ref{Recoil} we
sketch the calculation of the brane recoil velocity according to
the prescriptions of Ref. \cite{Ellis:1999uh,Ellis:1999jf,%
Ellis:1999sd,Ellis:2000sx,%
Gravanis:2002ii,Ellis:1999sf}. Then (Section
\ref{Sec:Dirac}) we analyze the modified Dirac equation according
to   the viewpoint (ii) above. Finally (Section
\ref{Sec:Concl}) we  consider some relevant experimental
results which are  compared with the theory's expectations and
then state our conclusions.

\section{Brane Recoil Velocity}
\label{Recoil}

 In order to have a quantitative description of the collision process
which satisfies the requirements of the theory in Ref
\cite{Ellis:2002bu}, which are (in the frame where the brane is
initially at rest)
  \begin{itemize}
  \item brane recoil velocity parallel to the momentum $\Vec{p}$ of
  the incoming   particle.
  \item Magnitude of the recoil velocity of order $\mid \Vec{p} \mid /
  M$,
  \end{itemize}
we consider a 1-dimensional description of the collision where the
 modeled as a large reflecting mirror of mass $M$ where the incoming
 particle is perpendicularly reflected.

The brane recoil velocity $\;\Vec{V}$ is defined by
\begin{equation}
\Vec{V}\equiv \frac{1}{M}\left( \Vec{p}-\Vec{r}\right),
\end{equation}
which we covariantize as
\begin{equation}
V^{\mu }=\frac{1}{M}\left( p^{\mu }-r^{\mu }\right)
=(V^{0},\;\Vec{V}\;),
\end{equation}
where $p^{\mu },\;r^{\mu }$ are the standard four momenta of the
in and out   ordinary particles, respectively.  We write
$p^{\mu }=(E,\;\Vec{p})$  use the standard
dispersion relation $E = \sqrt{p^2 + m^2}$ ($c=1$) together with the
notation:
\begin{equation}
\begin{array}{cc}
\Vec{p} = p\,\Vec{\hat{k}},& \Vec{r} = r\,\Vec{\hat{k}},\\ \Vec{w}
= w\,\Vec{\hat{k}},&  \Vec{u} = u\,\Vec{\hat{k}},
\end{array}
\end{equation}
 for the incoming and outcoming  particle's momenta and  D-brane
 velocities
respectively. Using  energy-momentum conservation in the
non-relativistic limit,  we obtain 
\begin{equation}
  V = 2\frac{p - mw}{M + m} = V(w),\quad \Vec{V} = V \Vec{\hat{k}} 
\end{equation}
which reduces to 
\begin{equation}
 V = \frac{2}{M}(p - mw) + O(M^{-2}) 
\end{equation}
 in the
large D-particle mass limit.

In the same way, the zeroth component of the recoil velocity 
(to order $1/M$) is just  
\begin{equation}
  V^0 = \frac{2w}{M} (p - m w) = wV.
\end{equation}
 The above equation is the relativistic statement that 
\begin{equation}
W^\mu V_\mu = 0 \label{eq:V.W}
\end{equation}
 which follows from the fact that $V^\mu $   is just
$\delta W^{\mu}$ while $W^{\mu} W_{\mu} $ is constant (i.e.$= 1$).
Thus to order $O(1/M)$ we have 
\begin{equation}
  V_{LAB} = \gamma_w V_{CMB}.
\end{equation}

 Coming back to vector notation, the above relations  take the form
\begin{equation}
  \begin{array}{cc}
  \Vec{V} = \frac{2}{M}(\Vec{p} - m\Vec{w}), & 
  V^0 = \Vec{w} \cdot \Vec{V}.
  \end{array}
\end{equation}

It is easily seen that $\Vec{V}$ is Galilean invariant to first
order in $w$, and is thus equal to $\Vec{V}_{CMB}$ (to such
order), 
 so we
introduce the notation
\begin{equation}
   \Vec{v} = \Vec{V}_{CMB} =  \frac{2}{M}\Vec{p}_{CMB} =
 \frac{2}{M}(\Vec{p}_{LAB} - m\Vec{w}) = \Vec{V}_{LAB}
                             \label{Def:v}
\end{equation}
 for the Galilean-invariant D-particle
recoil velocity.

 However,  $V^0$ is not Galilean invariant
\begin{equation}
  V^0_{LAB} =  \Vec{w} \cdot \Vec{v} \neq V^0_{CMB} = 0
                              \label{Cond:V0}
\end{equation}
and it is invariant  only under  the subset  of Galilean
transformations involving relative velocities orthogonal to the
D-particle recoil velocity. But nothing in the actual laboratory
experiments we will consider is restrictive in this way.

Equation (\ref{Def:v}) is a particular case of the condition 
considered in Ref \cite{Ellis:1999sf} as defining ``admissible''
Lorentz transformations: they should keep the magnitude of the
recoil velocity invariant: $\mid \Vec{V}_{CMB} \mid = \mid
\Vec{V}_{LAB} \mid$. The simple Galileo transformation leaves both
magnitude and direction of the vector invariant. 

\section{Reduction of the modified Dirac equation}
\label{Sec:Dirac}

We assume that equation (\ref{DIREQ}) describes the dynamics in
the CMB frame. In order to obtain the corresponding description in
the laboratory frame we rewrite (\ref{DIREQ}) in a covariant way
by introducing the laboratory CMB 4-velocity as $W^{\mu }=\gamma
\left( 1,\Vec{w} /c\right)$. Here we are imposing covariance under
observer Lorentz transformations, while we are violating particle
Lorentz invariance, in the notation of Ref. \cite{Colladay:1998fq}.

Let us consider the correction term
\begin{equation}
\Delta D= - {\gamma}^{0}i\,v^{k}\partial _{k}\,\Psi.
\end{equation}
The covariant extensions are
\begin{eqnarray}
 {\gamma}^{0} &\longrightarrow& {\gamma}^{\mu }W_{\mu }, \nonumber\\
  v^{k}\partial _{k} &\longrightarrow& V^{\mu }\left( g_{\mu \nu
  }-W_{\mu }W_{\nu }\right) \partial ^{\nu },
\end{eqnarray}
 which produce
\begin{equation}
 \Delta D=-\left(
{\gamma}^{\rho}W_{\rho}\right) V^{\mu }\left( g_{\mu \nu }-W_{\mu
}W_{\nu }\right) \partial ^{\nu }\,\Psi.
\end{equation}
This amounts to apply an standard Lorentz boost from the CMB frame
to the LAB frame.

  Here $V^\mu$ is the recoil
four velocity of the D-particle with respect to the laboratory
frame. The resulting Dirac equation is then
\begin{equation}
\left\{\gamma ^{\mu }i\,\partial _{\mu } - \gamma ^{\rho
}\left[W_{\rho }V^{\mu }\left( \delta _{\mu }^{\nu }-W_{\mu
}W^{\nu }\right)\right] i\,\partial _{\nu } -m \right\}\Psi =0.
\end{equation}

 From
 (\ref{eq:V.W}) the above equation simplifies to
\begin{equation}
 \left\{\gamma ^{\mu }i\,\partial _{\mu } - \gamma ^{\rho
}\left[W_{\rho }V^{\nu }\right] i\,\partial _{\nu } -m
\right\}\Psi =0,
                            \label{eq:DiracMod}
\end{equation}
which can be put in the  standard form of 
\cite{Kostelecky:1999zh,Colladay:2002eh}. The
identification of  the  corresponding Kosteleck\'y parameters
leads to
\begin{eqnarray}
\label{DEFLANE} a_{\mu } &=&b_{\mu }=0=H_{\mu \nu }, \qquad
g_{\lambda \mu \nu } = 0=d_{\mu \nu }, \qquad c_{\mu \nu } = -
W_{\mu }V_{\nu }.
\end{eqnarray}
Let us note that $c_{\mu\nu}$ is not symmetrical in this case.  Note
also that we have $c_{\mu\nu}\sim W_\mu V^{\rho} C_{\rho\nu}$, where
$C_{\rho\nu}$ is the analogous factor introduced in
Ref. \cite{Sudarsky:2002ue}. This makes explicit the difference
between the theories considered there and in Ref. \cite{Ellis:2002bu}.

According to \cite{Kostelecky:1999zh,Colladay:2002eh} the
corresponding Hamiltonian in the non-relativistic limit is
\begin{equation}
\label{NRHAM} h=m + \frac{p^2}{2m} - m c_{00} +
m(c_{0j}+c_{j0})\frac{p_j}{m}-m(c_{jk}+\frac{1}{2}c_{00}\delta_{jk})\frac{p_j
p_k}{m^2}
\end{equation}
up to fourth order in the momenta. Here $p_j$ are the components
of $- \Vec{p}$. Using (\ref{DEFLANE})  we find
\begin{equation}
\begin{array}{cc}
c_{00} = - \Vec{w}\cdot\Vec{v},
& c_{0\,i}=v^i\left(1 + \frac{w^2}{2}\right),  \\
c_{i\,0}=({\Vec v}\cdot{\Vec w})w^i, & c_{ij} = - w^{i}\,v^{j},
\end{array}
\label{COEFC}
\end{equation}
in the non-relativistic approximation, where we have kept terms up
to second order in ${\bf w}$. In this way, (\ref{NRHAM})
reduces to
\begin{equation}
 h=m\left( 1+{\Vec w}\cdot{\Vec v} \right) 
  - \left(1 + \frac{w^2}{2}\right) {\Vec p}\cdot{\Vec v} 
  + \left(1+{\Vec w}\cdot{\Vec v} \right)\frac{p^2}{2m} 
  + \frac{1}{m} \left[\Vec{v} \cdot (\Vec{p} - m \Vec{w})\right] \,
  (\Vec{w} \cdot \Vec{p}) 
  \label{Ham-Eff}
\end{equation}

Let us emphasize that the above Hamiltonian describes the
non-relativistic, linear in ${\Vec v}$ (the D-particle's recoil
velocity in the CMB frame) corrections to the spin 1/2 particle
dynamics in the laboratory frame. 
As it will be apparent in the sequel, the term $-\Vec{p} \cdot
\Vec{v}$ effectively contributes to the term $\frac{p^2}{2 m}$, which is
not relevant for clock comparison experiments.

Substituting (\ref{Def:v}) in Eq. (\ref{Ham-Eff}) the final
Hamiltonian is
\begin{equation}
h=m\left( 1 - \frac{2m}{M}{\Vec{w}}\,{}^{2}\right) +\left(
1-\frac{4m}{M} - \frac{20}{3} \frac{m}{M}{\Vec{w}}\,{}^{2}\right)
\frac{{\Vec p}\,{}^{2}}{2m} - 4
\frac{m}{M}\,\frac{w^{i}\,Q_{Pij}\,w^{j}}{m}, \label{EFFH}
\end{equation}
where $Q_{Pij} = p_i p_j - \delta_{ij} p^2/3$.  Here we have neglected
the terms linear and cubic in ${\Vec p}$ since the experiments in
question deal with a bound state, for which such  averages  vanish.

The corrections in (\ref{EFFH}) have the same form as those discussed
in Ref. \cite{Sudarsky:2002ue} and we only focus upon the two terms
which are probed by the measurements in clock comparison experiments:
the velocity-spin coupling and the quadrupolar correction to the
inertial mass. The first correction is absent in the D-particle recoil
model. In the notation of Ref. \cite{Sudarsky:2002ue} this leads to
\begin{equation}
\Theta_2+\frac{1}{2}\Theta_4=0
\end{equation}
and the model in fact evades the stringent limit settled in that
reference.  Next we concentrate on the quadrupolar mass correction
 $ \delta h_Q= -4 w^{i}\,Q_{Pij}\,w^{j}/M $.
 Comparison with the
corresponding term in Eq. (6) of Ref. \cite{Sudarsky:2002ue} leads to
the identification
\begin{equation}
\frac{4}{M}=\,\mid \Theta_1\,\ell_P \mid.
\end{equation}

This differs from the result in Ref. \cite{Ellis:2002bu}, which may be
due to our use of standard Observer Lorentz transformations, instead
of the modified ones proposed in reference \cite{Ellis:1999sf}.
However, as we have shown in Section \ref{Recoil}, a Galilean
transformation is a particular case of the latter, since it leaves
invariant both magnitude and direction of $\Vec{v}$ to lowest order.
 Yet the source of our effect is tied to the lack of invariance of
 $V^0$ as shown in Eq. (\ref{Cond:V0}).

Moreover, we have included all terms $O(1/M)$ and $O(w^2)$.
Indeed, the sensitivity of modern observational techniques is so
high that these terms, which were ignored in Ref. \cite{Ellis:2002bu},
are precisely those which are testable in clock comparison
experiments.  Our approach, based on standard Lorentz
transformations and the corresponding conservation laws, leads to
their presence in the non-relativistic Hamiltonian (\ref{EFFH}).

 In fact, it seems very
difficult to conceive  a scenario in which dispersions relations
could break Lorentz invariance while an effective isotropy is
preserved, i.e.  physics remains independent of the direction of
$\Vec{w}$, in frames moving with respect to the CMB frame.  The
point is that observer standard Lorentz transformations will map
such dispersion relations into direction-dependent terms in the
Hamiltonian appropriate for the local laboratory frame.

\section{Conclusions}
\label{Sec:Concl}

The bound we have previously found for $\Theta_1$ in Ref.
\cite{Sudarsky:2002ue} leads to
\begin{equation}
\label{BOUNDM} M\,> 1.2\times 10^5\, M_P,
\end{equation}
where $M_P$ is the Planck mass. This is, so far, the strongest bound
on the mass of the D-particle.  Let us note that this value for $M$ is
larger by several orders of magnitude than the values testable in the
gamma ray bursts experiments, given their expected sensitivity. The
latest astrophysical limit is $M \sim 6.9 \times 10^{15}$ GeV $\sim
5.6\times 10^{-4} M_P$ \cite{Ellis:2002in}, which is $\sim 10^{-8}$
times weaker than Eq. (\ref{BOUNDM}).

After averaging over the momenta in the bound state, the effective
D-particle recoil velocity is
\begin{equation}
{\Vec v} \rightarrow -2 \frac{m}{M}{\Vec w}, \label{RECVEL2}
\end{equation}
as can be seen from Eq. (\ref{Def:v}). Using also  Eq.
(\ref{BOUNDM}) we obtain  the  bound
\begin{equation}
\frac{\mid{\Vec v}\mid}{c}\,<\, 2 \times 10^{-27}, \label{Bound:v}
\end{equation}
for the  D-particle recoil velocity after colliding with
 a $m\approx 1$ GeV target  moving with the Earth
through the cosmic background.

  Let us remark that the results of this paper apply to a larger set
of quantum gravity models than that of reference
\cite{Ellis:2002bu}. Indeed, let there be a model having a
``privileged'' vector $U^\mu$ such that the Dirac equation takes the
form:
  \begin{equation}
   \left\{\gamma ^{\mu }i\,\partial _{\mu } - \gamma ^{\rho
}\left[W_{\rho }U^{\mu }\left( \delta _{\mu }^{\nu }-W_{\mu
}W^{\nu }\right)\right] i\,\partial _{\nu } -m \right\}\Psi =0. 
  \end{equation}
  Then, the reader can verify that, provided that the vector $U^\mu$
has the nonrelativistic limit $U^\mu \approx (1, \Vec{u})$, the
projected vector $\tilde{V}^\mu = U^\nu \left(\delta^\mu_\nu - W^\mu
W_\nu \right)$ with components $\tilde{V}^\mu = (\Vec{w} \cdot
\Vec{u}_{CMB}, \Vec{u}_{CMB})$ in the non-relativistic limit, has the
same properties of $V^\mu$ and will satisfy a similar bound to
(\ref{Bound:v}) provided $\Vec{u}_{CMB} \propto \Vec{p}_{CMB}$.

Our analysis shows that the existing high precision experimental
results are able to set very stringent bounds on the mass and the
recoil velocity of the D-particle. The latter bound indicates that
such velocity is extremely small, even by the standards of every
day  experience (e.g.  the speed of a crawling snail is
$(\sim 10^{-11} $c$)$). Although it is in principle conceivable
that further refinements could possibly be used to detect the
recoil velocity of the D-particle, this last observation suggests
that it would be quite unlikely that such effects can ever be
detected.

We have found that after identifying the preferred frame of
reference associated with Planck scale physics effects with the
CMB frame, existing results of atomic and nuclear physics
experiments can be translated into very strict bounds on the
quantum gravity induced modifications to the propagation of Dirac
fields. The present results indicate that using the type of
experiments we have considered, string theory inspired schemes are
as testable as those in loop quantum gravity. Moreover, they seem
to cast doubts on the possibility to detect such effects in the
near future, using the type of experiments with high energy gamma
bursts which have been advocated in
\cite{Amelino-Camelia:1998gz,Ahluwalia:1999aj,%
Amelino-Camelia:1999zc}.

\begin{acknowledgments}
We wish to ackowledge the authors of Ref. \cite{Ellis:2002bu} for very
valuable comments and several other colleagues for criticisms.
D.S. would like to acknowledge partial support from DGAPA--UNAM
Project No IN 112401, a CONACYT sabbatical fellowship, and the
Eberly Endowment and thanks Penn State University for its
hospitality. L.U acknowledges partial support from DGAPA--UNAM
Project No IN-117000 and CONACYT project 40745-F. H. V.
acknowledges a leave of absence from Universidad Nacional de La Plata.
\end{acknowledgments}

\bibliography{QGRTest}

\end{document}